\documentclass[conference]{IEEEtran}
\IEEEoverridecommandlockouts
\usepackage{tabularx}
\usepackage{algorithm}
\usepackage{algpseudocode}
\usepackage{cite}
\usepackage{amsmath,amssymb,amsfonts}
\usepackage{graphicx}
\usepackage{textcomp}
\usepackage{algpseudocode}
\usepackage{xcolor}
\def\BibTeX{{\rm B\kern-.05em{\sc i\kern-.025em b}\kern-.08em
    T\kern-.1667em\lower.7ex\hbox{E}\kern-.125emX}}
\begin{document}
\newcommand{\shah}[1]{{\color{red}#1}}
\newcommand{\brian}[1]{{\color{blue}#1}}
\newcommand{\naveen}[1]{{\color{cyan}#1}}

\title{
Experimental Study of Adversarial Attacks on ML-based xApps in O-RAN


}

\author{\IEEEauthorblockN{Naveen Naik Sapavath\IEEEauthorrefmark{1}\textsuperscript{\textsection},
Brian Kim\IEEEauthorrefmark{2}\textsuperscript{\textsection},
Kaushik Chowdhury\IEEEauthorrefmark{2}, and 
Vijay K Shah\IEEEauthorrefmark{1}}
\IEEEauthorblockA{\IEEEauthorrefmark{1}NextG Wireless Lab,
George Mason University, Fairfax, VA 22030 USA}
\IEEEauthorblockA{\IEEEauthorrefmark{2}\textit{Institute for the Wireless Internet of Things}, Northeastern University, Boston, USA}

}

\maketitle

\begingroup\renewcommand\thefootnote{\textsection}
\footnotetext{Equal contribution. This work was partially supported by the NSF under award \#2120411, and in part by Commonwealth Cyber Initiative (CCI), an investment by the Commonwealth of Virginia in the advancement of cyber R\&D, innovation, and workforce development. The authors gratefully acknowledge the funding from the US National Science Foundation (grant CNS-2112471)}
\endgroup

\begin{abstract}
Open Radio Access Network (O-RAN) is considered as a major step in the evolution of next-generation cellular networks given its support for open interfaces and utilization of artificial intelligence (AI) into the deployment, operation, and maintenance of RAN. However, due to the openness of the O-RAN architecture, such AI models are inherently vulnerable to various adversarial machine learning (ML) attacks, i.e., adversarial attacks which correspond to slight manipulation of the input to the ML model. In this work, we showcase the vulnerability of an example ML model used in O-RAN, and experimentally deploy it in the near-real time (near-RT) RAN intelligent controller (RIC). Our ML-based interference classifier xApp (extensible application in near-RT RIC) tries to classify the type of interference to mitigate the interference effect on the O-RAN system. We demonstrate the first-ever scenario of how such an xApp can be impacted through an adversarial attack by manipulating the data stored in a shared database inside the near-RT RIC. Through a rigorous performance analysis deployed on a laboratory O-RAN testbed, we evaluate the performance in terms of capacity and the prediction accuracy of the interference classifier xApp using both clean and perturbed data. We show that even small adversarial attacks can significantly decrease the accuracy of ML application in near-RT RIC, which can directly impact the performance of the entire O-RAN deployment.


\end{abstract}



\section{Introduction}

The Open Radio Access Networks (O-RAN) framework aims to transform 5G and beyond by exploiting open, virtualized, and fully interoperable radio access networks (RANs). O-RAN has many advantages compared to the traditional RAN architecture due to disaggregated and virtualized components including open-source software elements from different vendors connected through open interfaces \cite{OranWG1}. Also, O-RAN architecture integrates a modular base station software stack on off-the-shelf hardware allowing baseband and radio unit components from different providers to operate smoothly. 

The extensive reconfigurability made possible via O-RAN makes it suitable for machine learning (ML)-based network control. Indeed different ML methods, include deep learning (DL), have been shown to be effective in 5G and beyond (NextG) applications, ranging from user equipment (UE) identification for initial access, MIMO beam optimization, and resource allocation for network slicing. Despite the many benefits of open interfaces, softwarization and virtualization in O-RAN open the door for exploitation through advanced algorithmic techniques, including ML, which can be used at third-party microservices, known as xApps and rApps hosted by near-real time (near-RT) and non-real time (non-RT) RAN Intelligent Controller (RIC) respectively in O-RAN. The RICs provide the ability for ML to optimize network performance and user experience~\cite{lee2020hosting, bonati2021intelligence,niknam2022intelligent}. 

\begin{figure}[t!]
\centering
\includegraphics[width = 0.37\textwidth]{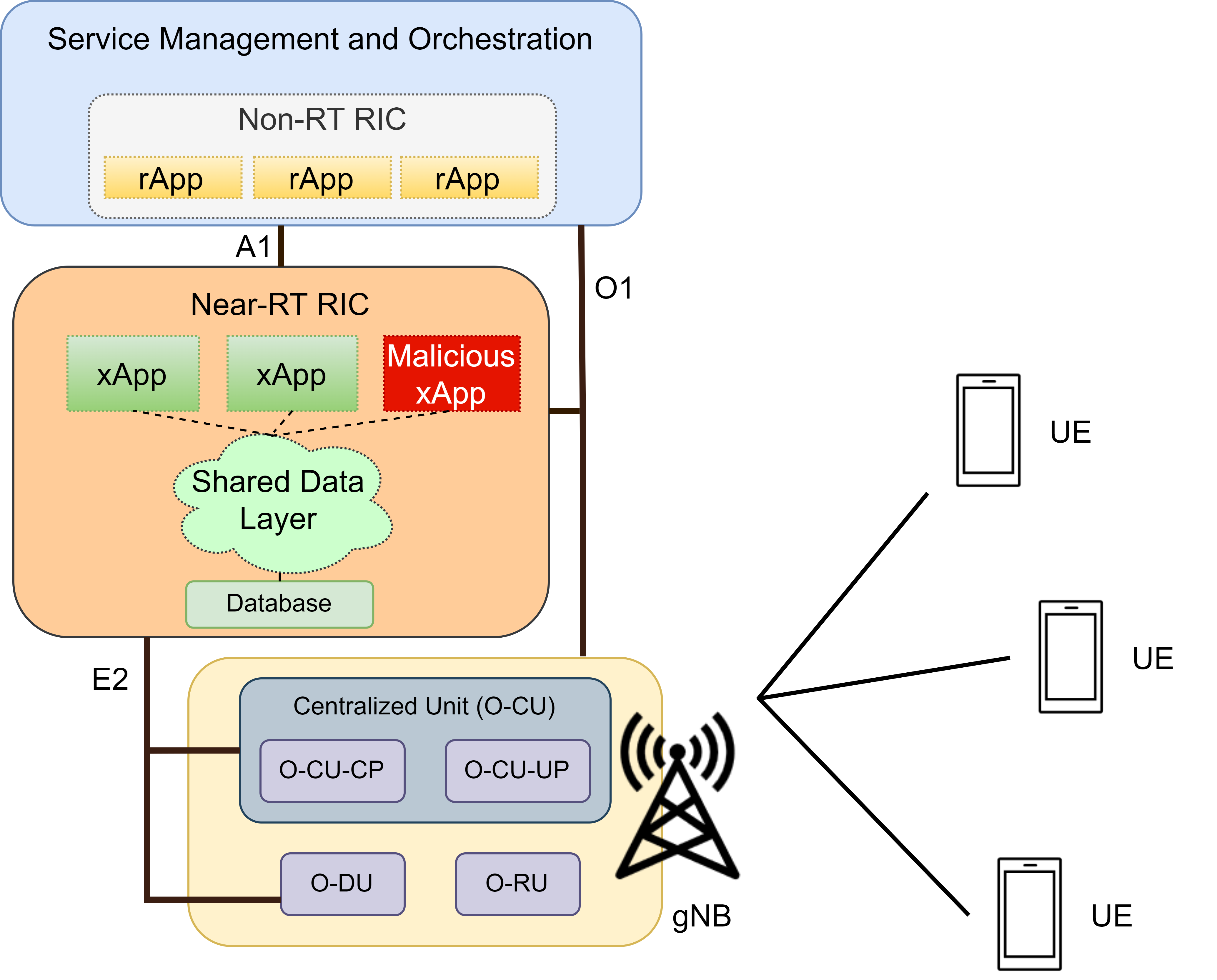}
\caption{Overview of O-RAN architecture with components and interfaces including malicious xApp in near-RT RIC.} \label{fig:ORAN system}

\end{figure}

\noindent$\bullet$ \textbf{ML-related Vulnerabilities:} Deep neural networks (DNNs) is a popular branch of ML, which minimizes the need for domain knowledge for inference tasks. While it has seen great success in the field of computer vision, DNNs have also been used earlier for RF fingerprinting \cite{jian2020deep}, modulation classification \cite{soltani2019spectrum}, beam selection \cite{salehi2022deep}, among others. However, DNNs are  also vulnerable to attacks that are being actively studied in the emerging field of adversarial machine learning \cite{Vorobeychik1}. Since, the shared and broadcast nature of the wireless medium increases the potential for adversaries to manipulate DL-based wireless applications,   attacks such as inference attacks \cite{shi2018adversarial}, poisoning attacks \cite{sagduyu2019adversarial}, and evasion attacks or adversarial attacks \cite{kim2021channel} have been identified. The  RICs that execute  ML-based xApps are also susceptible to these attacks, due to the open and shared nature of the O-RAN~\cite{liyanage2023open}. However, to the best of our knowledge, there has been no work that studies the vulnerability of the ML-based xApp deployed at near-RT RIC against adversarial attacks and its impact on the network performance of O-RAN.


\noindent$\bullet$ \textbf{Study of  Vulnerabilities in ORAN:} In this paper, we systematically explore the vulnerability of the O-RAN framework, especially the DNN-based xApp used in near-RT RIC due to adversarial attacks. As a case study for the vulnerability of an xApp, we consider an xApp that classifies the wireless interference type \cite{schmidt2017wireless,grunau2018multi} received at the base station (gNB) which has been studied in the wireless communication field since a signal of interest can be easily disrupted by another signal as they share the same radio band. We deploy an interference classifier xApp into the O-RAN-compliant framework and design adversarial attacks which aim to slightly manipulate the original input that in turn forces misclassification. 

\noindent$\bullet$ \textbf{Contributions:}
We conduct an experimental campaign with an actual O-RAN deployed in the laboratory environment to analysis how a simple adversarial attack that manipulates data consumed by an xApp can largely affect the network performance.  In summary, our contributions are:
\begin{itemize}
    \item We assess the vulnerability of the xApp, we deploy pre-trained DNN-based xApp for interference classification at near-RT RIC in an O-RAN-compliant framework. 
    \item We investigate how adversarial attacks from malicious xApp from third-party vendors can deteriorate the performance of the legitimate xApp in the near-RT RIC.
    \item Utilizing the in-lab O-RAN testbed, we evaluate the impact of the adversarial attack on the ML-based interference classifier xApp which has been trained using a publicly available RFI dataset.  Furthermore, we utilize a synthetic 5G dataset for analyzing the network performance of O-RAN deployments when the ML-based xApp is affected by the adversarial attack. Our analysis shows that even small adversarial attacks significantly compromise the accuracy of ML-based xApp, which causes significant degradation of the network performance (measured as average capacity and the number of bits lost). 
    \item We pledge to release the source code, dataset, and open source O-RAN implementation code to the community for further research.
\end{itemize}

\section{O-RAN Background}
Within the O-RAN architecture shown in Fig. \ref{fig:ORAN system},  we mainly focus on two important components in this paper.
\subsubsection{\textbf{RIC}}
RAN Intelligent Controller (RIC) is a key component of O-RAN which is a software-defined subsystem that optimizes the RAN functions and controls different units in the open access network through interfaces \cite{bonati2023openran}. Two RICs, non-RT RIC and near-RT RIC, play an important role in supporting O-RAN's capability of network function disaggregation and smooth operations among different applications from third-party vendors, which is also known as multi-vendor interoperability. These third-party vendors develop various microservice-based applications, known as xApps and rApps deployed in non-RT RIC and near-RT RIC respectively, to automate and optimize RAN operations such as policy scheduling \cite{bonati2021intelligence}. AI/ML-driven applications have been used in RICs to greatly reduce the mobile operator's total cost and also enhance the quality of experience for the users. 

 $\bullet$ \textbf{Non-RT RIC:} 
Non-RT RIC performs all its operations and tasks within the automation platform called Service Management and Orchestration (SMO) where the non-RT RIC is in charge of tasks on a time scale longer than 1 second. Non-RT RIC has the capability of controlling RAN elements and their resources which gives the ability to indirectly supervise all the components that are connected to the SMO.


$\bullet$ \textbf{Near-RT RIC:} Near-RT RIC is located at the edge of the network which communicates with the Central Unit (CU) and Distributed Unit (DU) through the E2 interface as we can see from Fig. \ref{fig:ORAN system}. Near-RT RIC operates within a 10ms and 1s time frame where it consists of multiple microservice applications called xApp, a shared database layer (SDL) containing information on the RAN, and messaging infrastructure to communicate with different components in the systems. Through A1 and O1 interfaces, near-RT RIC communicates with non-RT RIC for optimization and task management across the RAN using the real-time RAN data.


\subsubsection{\textbf{xApp}}
 xApp is a plug-and-play microservice-based software application deployed in near-RT RIC to implement specific functions or services. Within the near-RT RIC, xApps, which can be developed by a third party, access the data in SDL for resource management, RAN data analysis, and RAN control. After accessing the data, xApp sends back control decisions using the E2 interface.
 


\section{System and Attack Model}\label{sec: System and attack model}
We consider an O-RAN system where there exists a legitimate xApp in the near-RT RIC that sends control actions to gNB by using the received data from the RAN. We assume that legitimate xApp can access data stored in the database through SDL to make a decision in near-RT RIC. However, due to the openness of the O-RAN framework and the structure of the near-RT RIC, there exist numerous potential security threats.


In this paper, we focus on the security problem of O-RAN using an example adversarial attack involving  small modifications of the original input to the DNN such that the DL algorithm misclassifies the input. Through this approach, we show the vulnerability of an xApp in the near-RT RIC against adversarial attacks. In this case, the attacker is a \textit{malicious} xApp, deployed by a third-party vendor as seen in Fig. \ref{fig:ORAN system}. This xApp carefully perturbs the original input with slight changes and stores it in the database through SDL so that the legitimate xApp returns incorrect classification using such altered data. We demonstrate two different well-known adversarial attacks called (i) the fast gradient sign method (FGSM) \cite{kurakin2016adversarial} and (ii) projected gradient descent (PGD) attack \cite{madry2017towards} that distorts the original data in the database.

\subsection{FGSM attack}
The FGSM is a one-step attack based on the gradient that can generate an adversarial attack in a computationally efficient manner. FGSM creates adversarial attack $\boldsymbol{x}_{adv}$ that maximizes the loss function $\mathcal{L}(\boldsymbol{\theta},\boldsymbol{x},\boldsymbol{y})$ by bounding $L_{\infty}$-norm where $\boldsymbol{\theta}$ is the model weights, $\boldsymbol{x}$ is the input to the DNN, and $\boldsymbol{y}$ is the corresponding label. The adversarial attack \cite{kurakin2018adversarial} can be expressed as $ \boldsymbol{\delta} = \epsilon \text{sign}(\nabla_{\boldsymbol{x}}\mathcal{L}(\boldsymbol{\theta},\boldsymbol{x},\boldsymbol{y}))$, 
where $\nabla_{\boldsymbol{x}}\mathcal{L}(\boldsymbol{\theta},\boldsymbol{x},\boldsymbol{y})$ is the gradient of $\mathcal{L}(\boldsymbol{\theta},\boldsymbol{x},\boldsymbol{y})$ with respect to the input data $\boldsymbol{x}$ and $\epsilon$ is the power scaling factor of the attack. Then, the perturbed input is represented as $\boldsymbol{x}_{adv} = \boldsymbol{x} + \boldsymbol{\delta}$.

\subsection{PGD attack}
The FGSM can be improved by running a much more thorough optimization by running an iterative algorithm where this multi-step iterative algorithm based on FGSM is called PGD attack \cite{madry2017towards}. PGD attack performs FGSM with a small step size $\alpha$ and projects the manipulated input onto the $L_{\infty}$-ball around the original input. The iterative method for $N$-step PGD attack is defined for $n$-th iteration as follows
\begin{align}\label{eq:pgd attack}
    &\boldsymbol{x}_{0} = \boldsymbol{x} \nonumber\\
    &\boldsymbol{x}_{n} = \text{clip}_{[\boldsymbol{x},\epsilon]}\{\boldsymbol{x}_{n-1}+\alpha \text{sign}(\nabla_{\boldsymbol{x}_{n-1}}\mathcal{L}(\boldsymbol{\theta},\boldsymbol{x}_{n-1},\boldsymbol{y}))\} \nonumber\\ 
    &\boldsymbol{x}_{adv} = \boldsymbol{x}_{N},
\end{align}
where $\text{clip}_{[\boldsymbol{x},\epsilon]}(\cdot)$ is elementwise clipping to $[\boldsymbol{x}-\epsilon,\boldsymbol{x}+\epsilon]$ so that the outcome stays in $L_{\infty} \epsilon-$neighborhood of $\boldsymbol{x}$.



We consider a white-box attack meaning that the adversary has all the knowledge about the attack target. Note that while this white-box attack can be extended to a more realistic attack using limited information as done in \cite{kim2021channel}, the focus of this paper is to study the potential security problem of the DL-based xApp in O-RAN systems. To show the vulnerability of xApps against adversarial attacks, we consider the interference classifier xApp and use the radio frequency interference (RFI) dataset \cite{ujan2020efficient}. Further, to analyze the impact of the adversarial attack on network performance, we create our own RFI dataset which we discuss in the next section.

\section{Interference Classification Dataset} \label{sec: dataset}

We selected a publicly accessible wireless interference dataset \cite{ujan2020efficient} that contains scalograms of a signal of interest (SOI) to which three different interference types- chirp interference (CI), continuous-wave interference (CWI), and multi-continuous-wave interference (MCWI) are added. These interference signals are generated at various signal-to-interference and noise ratios (SINR), and SOI is a real-time video stream processed and modulated by GNU radio before being transmitted using a Universal Software Radio Peripheral (USRP-N210). The radio employs a combiner to mix the SOI with the three distinct interference signals. Since we have no information about the specification of the data, i.e., a specific level of SINR for each scalogram, it is extremely difficult to demonstrate the network performance of the system based on the scalogram with and without the attack. 
\begin{figure}[t]
\centering
\includegraphics[width=0.35\textwidth]{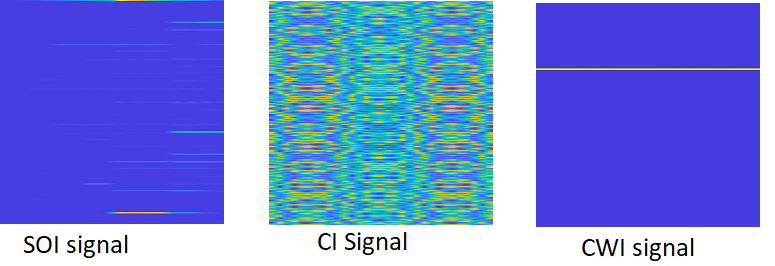}
\caption{Three different signal interference: SOI, CI, CWI.} \label{fig:Three}
\end{figure}

To demonstrate the network performance before and after the attack, we create a new dataset using MATLAB simulations in a 5G network setting. This was accomplished by assuming a specific number of users and channels. We generate 5G SOI with different center frequencies where we add a CI and CWI for each SOI. Note that we do not consider MCWI as the interference type for our synthetic dataset because this dataset has been created considering a realistic 5G network deployment where the gNB operates one unique channel of a prefixed channel bandwidth 200 MHz. After generating the three types of signals, which are SOI, CI, and CWI, we utilize the continuous wavelet transform (CWT) from the MATLAB toolbox to convert them into scalogram images as shown in Fig. \ref{fig:Three}. Note that we will call this dataset a synthetic dataset. CWT is a technique used to visualize In-phase and Quadrature (IQ) matrices by converting them into scalograms, which are scaled representations of the original signal's CWT.

To generate SOI in a 5G network setting, we have created a modulated signal using a specific modulation scheme, such as Quadrature Amplitude Modulation (QAM). We define the parameters of a QAM-modulated signal in MATLAB where we define $f_c = 3.5\times10^9$ Hz is the carrier frequency, $f_s = 10f_c$ is the sampling frequency, and signal-to-noise ratio (SNR) is SNR = 30 dB. We generate the modulated signal using the function called $qammod(signal, M)$ in Matlab, where $M$ represents the order of the Quadrature Amplitude Modulation (QAM) scheme used for modulating the input signal. To generate the different SOI signals, we use the following equation (\ref{SOI_our})\cite{proakis2008digital,haykin2008communication}. We employed the following equation (\ref{SOI_our}) to generate 4,000 distinct signals of interest (SOIs) with carrier frequencies spanning the range from 3.4 GHz to 3.6 GHz defined in the table \ref{tab:caption}:
\begin{equation}
\label{SOI_our}
SOI = real(\text{modulated signal}.* exp(2\pi f_{c}it)),
\end{equation}
where we have done the elementwise multiplication of the modulated signal data and the carrier signal to generate the SOI.  The modulated signal is the signal modulated using the QAM modulation scheme, $f_{c}$ is 5G carrier frequency, $i$ is the imaginary unit, and $t$ is a time vector. This process generates SOIs with unique frequencies within the specified range.
To generate a chirp signal in the sub-6 GHz range, we use the following formula \cite{proakis2008digital}
\begin{equation}
\label{CI_our}
CI = A_{CI}sin(2\pi t(f_{start}+0.5kt)),
\end{equation}
where $A_{CI}$ is the amplitude of the chirp signal, $f_{start}$ is the starting frequency of the chirp signal, $t$ is the time variable varies between $(0 \le t \le T)$, and $f_{end}$ is the ending frequency of the chirp signal. Finally, $k$ is the frequency sweep rate, which represents the rate at which the frequency changes over time where $k = (f_{end}-f_{start})/T$  so that the signal sweeps from $f_{end}$ to $f_{start}$, and $T$ is the sweeping duration.

Similarly, we have generated the CWI signal using the following equation  \cite{haykin2008communication}

\begin{equation}
\label{CWI_our}
CWI = A_{CWI}sin(2\pi(f_{cwi}t)),
\end{equation}
where $A_{CWI}$ and $f_{cwi}$ represent the amplitude and the frequency of the CWI signal, respectively. $f_{cwi}$  it is the frequency of the interfering sinusoidal signal that may affect the overall communication performance by adding noise or interference to the signal of interest.

\section{Case Study: Implementation of DL-based Interference Classification xApp}\label{sec:implementation}

\begin{figure}[t]
\centering
\includegraphics[width=0.48\textwidth]{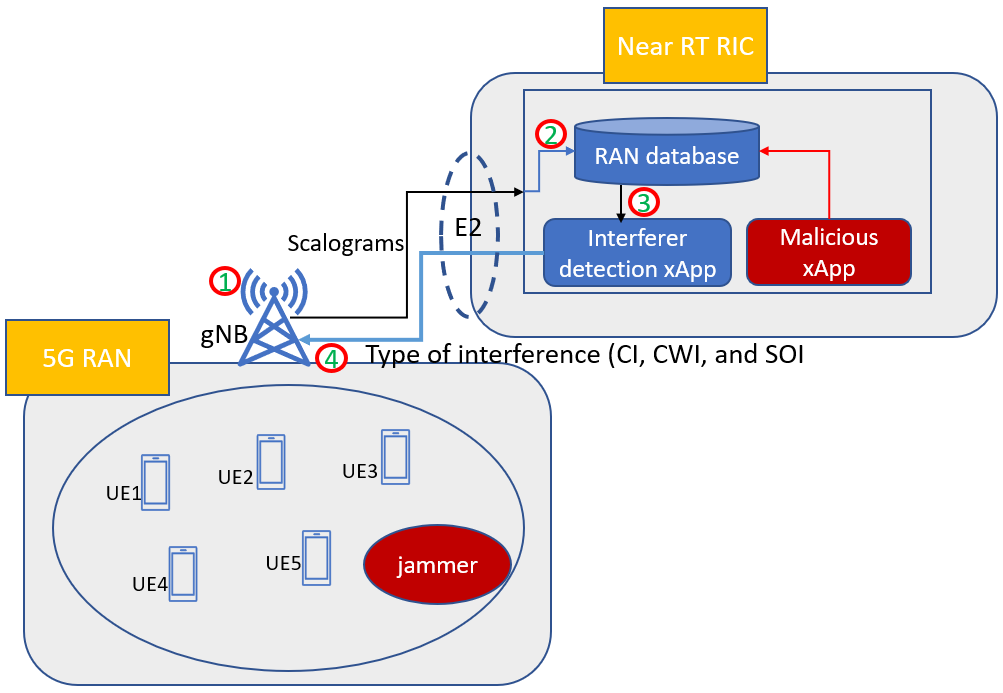}
\caption{Interference classification xApp deployed in the O-RAN based 5G network framework} \label{fig:ORAN based 5G network}

\end{figure}

In the O-RAN architecture depicted in Fig. \ref{fig:ORAN based 5G network}, we have implemented two components in O-RAN-compliant framework. First, we developed an interference classification xApp and deployed it in the near-RT RIC. Second, we utilized the MATLAB network simulator to simulate 5G network based on specific parameters shown in Table \ref{tab:caption}. 
The entire process of our interference classifier xApp implementation is outlined in four steps which are depicted as numbers in Fig. \ref{fig:ORAN based 5G network}.

\begin{table}[t]
\caption{5G parameters used in MATLAB simulation.}
\begin{tabular}{|c|c|}
\hline
 \textbf{5G parameters} & \textbf{value}  \\ \hline
 Considered frequency  & 3.5 GHz (C-band) - 3.4 GHz to 3.6 GHz  \\ \hline
 Transmit power  & 20 watts (43 dBm) per channel  \\ \hline
 SNR  & 0 to 30 dB  \\ \hline
Total allocated bandwidth  & 200 MHz  \\ \hline
 Number of channels  & 40  \\ \hline
 Number of users  & 50  \\ \hline
 Number of jammers  & 1 \\ \hline
\end{tabular}\label{tab:caption}
\end{table}

  

\textbf{Step 1:} To establish the connection between the gNB and an xApp, an xApp requests subscription to the gNB, and once subscribed, xApp is able to receive indication messages and send control messages, as specified in O-RAN's E2 specification. Indication messages are sent from the gNB to the xApp and can be used to transmit scalogram data, whereas control messages allow the xApp to make decisions and affect the behavior of the gNB.

\begin{algorithm}[t]
\caption{5G Network Interference Mitigation Algorithm}\label{alg:InterferenceMitigation}
\begin{algorithmic}
\Require xApp\_classification (CI, CWI, or SOI)
\Function{Process\_Interference}{xApp\_classification}
    
        \If{xApp\_classification is CI}
            \State Select a lower modulation coding scheme
        \ElsIf{xApp\_classification is CWI }
            \State Perform frequency hopping
        \ElsIf{xApp\_classification is SOI}
            \State gNB does not take any action

        \EndIf
\EndFunction
\end{algorithmic}
\end{algorithm}

\textbf{Step 2:} We have implemented the E2 interface and xApp uses a HTTP GET/POST application programming interface built upon a Flask server (SDL in the RIC of the O-RAN framework). Using the HTTP interface, the xApp and gNB are able to access and update a database stored in the Flask server, which is similar to O-RAN's SDL. 

\textbf{Step 3:} In our implementation, the MATLAB simulator submits scalograms to the Flask server, then the xApp downloads them from the server, makes a prediction, and updates the database with the prediction. 

\textbf{Step 4:} Finally, the MATLAB simulator gNB retrieves the prediction from the xApp where gNB applies mitigation technique, \textsc{Process\_Interference}, to reduce the impact of interference based on the received prediction. The details of the mitigation technique are described in Algorithm \ref{alg:InterferenceMitigation}.

Since we use the synthetic dataset which has two types of interference, we implement two mitigation techniques for each interference type. To combat CI, lower-order QAM schemes (QAM 32, 64, 128, and 256) are used, decreasing sensitivity to noise and interference. For CWI, the gNB employs frequency hopping or channel switching to maintain robust communication links and high-quality network performance.

After implementing the interference classifier xApp, we introduce adversarial attacks, as described in Section \ref{sec: System and attack model}, to the interference classifier xApp. In our implementation, there exists a malicious xApp that accesses the Flask server and perturbs the original data that is stored to cause misclassification at the interference classifier xApp. Note that when the interference classifier xApp misclassifies due to the adversarial attacks from the malicious xApp, the mitigation technique will also be applied incorrectly.

\section{O-RAN Performance Analysis}

During the performance analysis, we deploy a pre-trained interference classifier xApp into an O-RAN-compliant system, described in Section \ref{sec:implementation}, and analyze the effect of the adversarial attack. We use two datasets described in Section \ref{sec: dataset} where the real RFI dataset is used for analyzing the impact of the adversarial attack on interference classifier xApp and the synthetic dataset is used to measure the network performance when affected by the adversarial attack. We use Resnet18 \cite{he2016deep} followed by a fully-connected layer and softmax output layer for both datasets. For the RFI dataset, we use 3800 samples for training and 960 samples for testing. Note that the samples for each class (SOI, CI, CWI, MCWI) of interference are equally distributed for training and test dataset. We set the batch size as 256 and the number of epochs as 15 during the training. After the training, we obtain 98.5\% accuracy with the test dataset, as also seen in \cite{oyedare2023keep}. For the synthetic dataset, we generate a total number of 12000 samples where 4000 samples for each class (SOI, CI, CWI) are generated. We use 9600 samples during training and 2400 samples for testing which results in 99.9\% accuracy during testing. Note that we use $\alpha = 5$ and $N=20$ for the PGD attack in equation (\ref{eq:pgd attack}) for our evaluation.

\subsection{Performance of Interference classifier xApp}

\begin{figure}[t]
\centering
\includegraphics[width=0.41\textwidth]{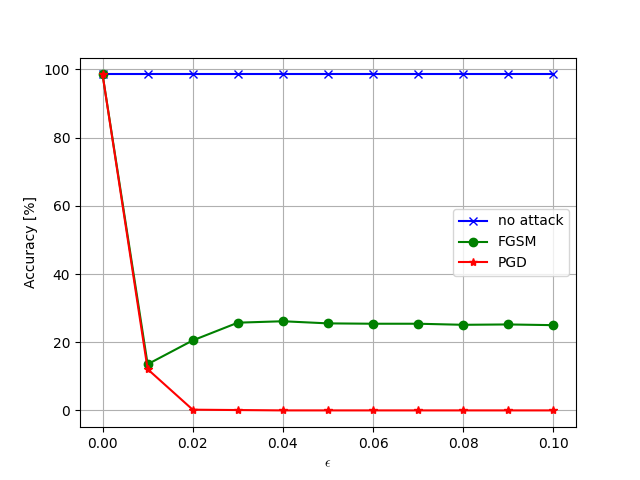}
\caption{Interference classifier xApp accuracy before and after different types of attacks using RFI dataset.} \label{fig:attack performance not our}
\end{figure}

First, we deploy an interference classifier xApp that is trained using RFI dataset to explore the impact of adversarial attack. As we can observe in Fig. \ref{fig:attack performance not our}, both FGSM and PGD adversarial attacks significantly decrease the performance of the DL-based xApp using real RFI data even though small $\epsilon$ is used to generate the attack. Note that $\epsilon$ denotes the power level of the attack where it represents the percentage of change that can be made to each pixel of an image. Moreover, it is seen that the PGD attack outperforms the FGSM attack as we expected since the PGD attack is generated using FGSM with the iterative method.

\begin{figure}[t]
\centering
\includegraphics[width=0.41\textwidth]{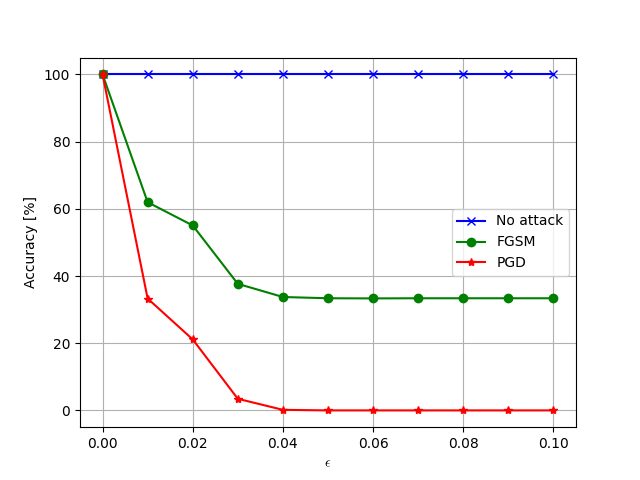} 
\caption{Interference xApp accuracy before and after different types of attacks using synthetic dataset.} \label{fig:attack performance our}
\end{figure}

We further investigate the performance of the adversarial attack by deploying the interference classifier xApp that is trained using a synthetic dataset in Fig. \ref{fig:attack performance our}. Similar to Fig. \ref{fig:attack performance not our}, we observe that interference classifier xApp is greatly affected by the adversarial attack. Specifically, even with $\epsilon=0.04$, we achieve zero accuracy when the PGD attack is used. Also, we observe that even with greater power used for FGSM attack the accuracy is not affected since FGSM attack is not carefully designed. Moreover, the FGSM attack causes random guessing for the interference classifier when high power is used, i.e., 25\% for Fig. \ref{fig:attack performance not our} and 33\% for Fig. \ref{fig:attack performance our}.


\subsection{Network Performance before and after attack}

\begin{figure}[t]

\centering
\includegraphics[width=0.45\textwidth]{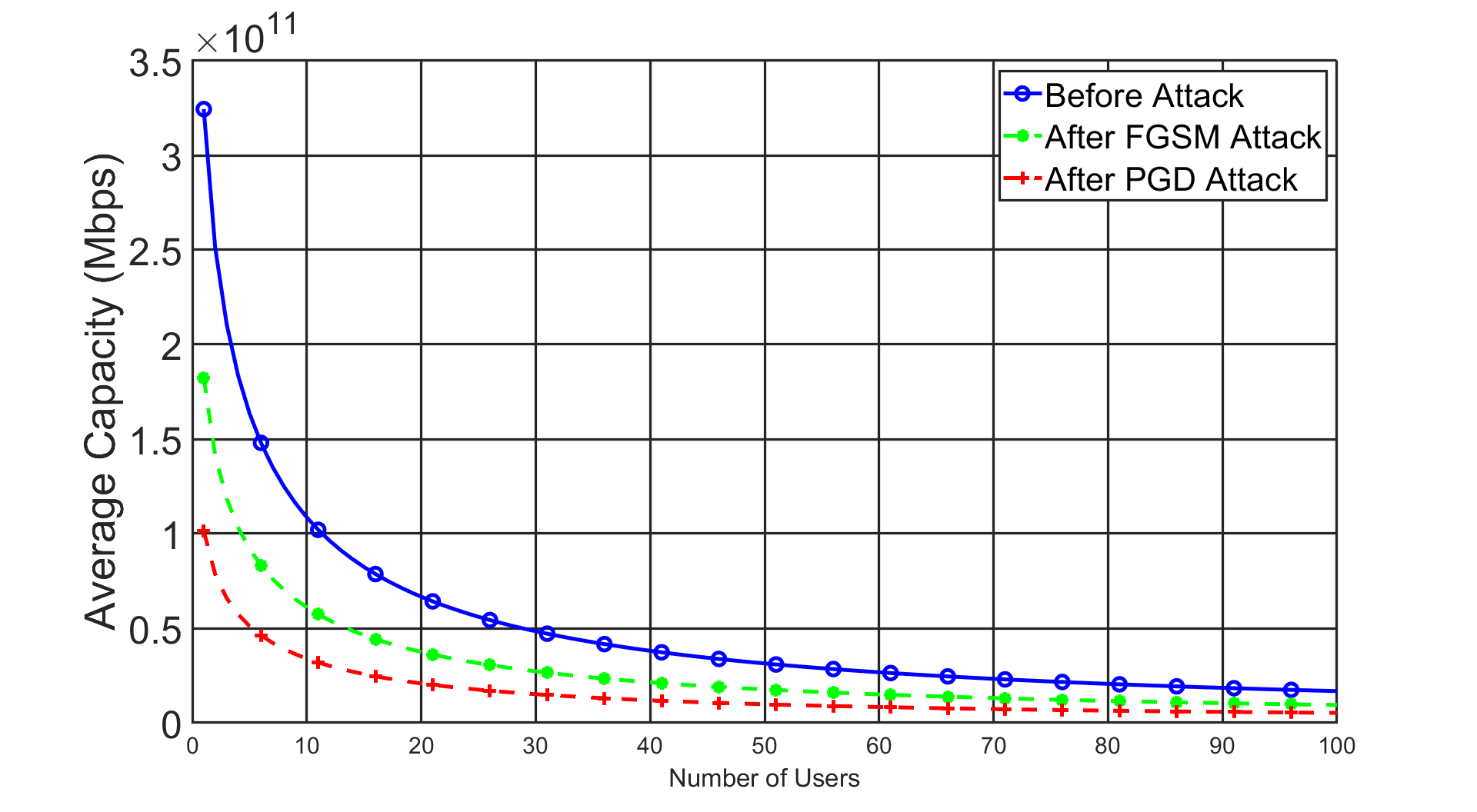}
\caption{Average capacity comparison of before attack and after the attack.} \label{fig:attack performance average capacity}
\end{figure}

Now, we assess the network performance when there exists malicious xApp to study its impact on the system. Note that we use FGSM and PGD attack with $\epsilon =0.05$ for network performance analysis. For Fig. \ref{fig:attack performance average capacity}, our evaluation metric is the overall network average capacity to capture the importance of classifying the interference. We consider a scenario with varying numbers of users from 1 to 100, where each user experiences satisfactory and stable network capacity without the malicious xApp. However, when there exists a malicious xApp, the legitimate xApp begins to misclassify signals, leading to a decrease in the average network capacity, as shown in Fig. \ref{fig:attack performance average capacity}. Specifically, we observe that the PGD attack greatly impacts the average capacity compared to the FGSM attack. 

\begin{figure}[t]
\centering
\includegraphics[width=0.45\textwidth]{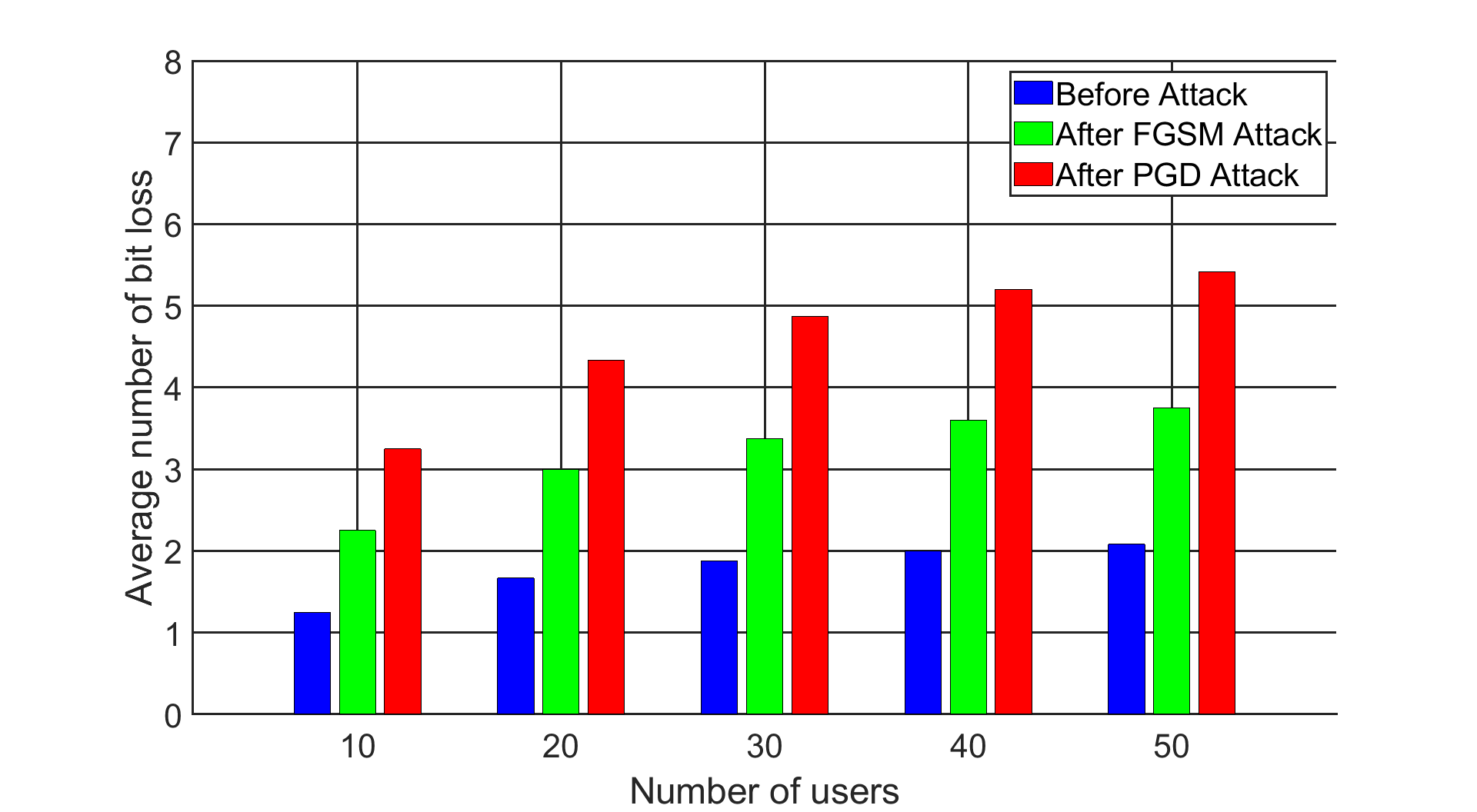}
\caption{Average number of bits loss per symbol before and after attack. } \label{fig:attack performance 1}


\end{figure}

We have evaluated network performance in terms of the average number of bit losses per symbol for varying numbers of users from 1 to 50 in Fig. \ref{fig:attack performance 1}. We observe that the xApp frequently misclassifies the interference as CI, resulting in a change in the MCS as the chosen mitigation technique. This change leads to a significant increase in bit losses. We observe that the bit loss before the attack is considerably lower compared to the post-attack scenario (both FGSM and PGD attacks), where a higher bit loss occurs due to the incorrect implementation of mitigation techniques at the gNB due to misclassification at the xApp.

\begin{figure}[t]
\centering
\includegraphics[width=0.5\textwidth]{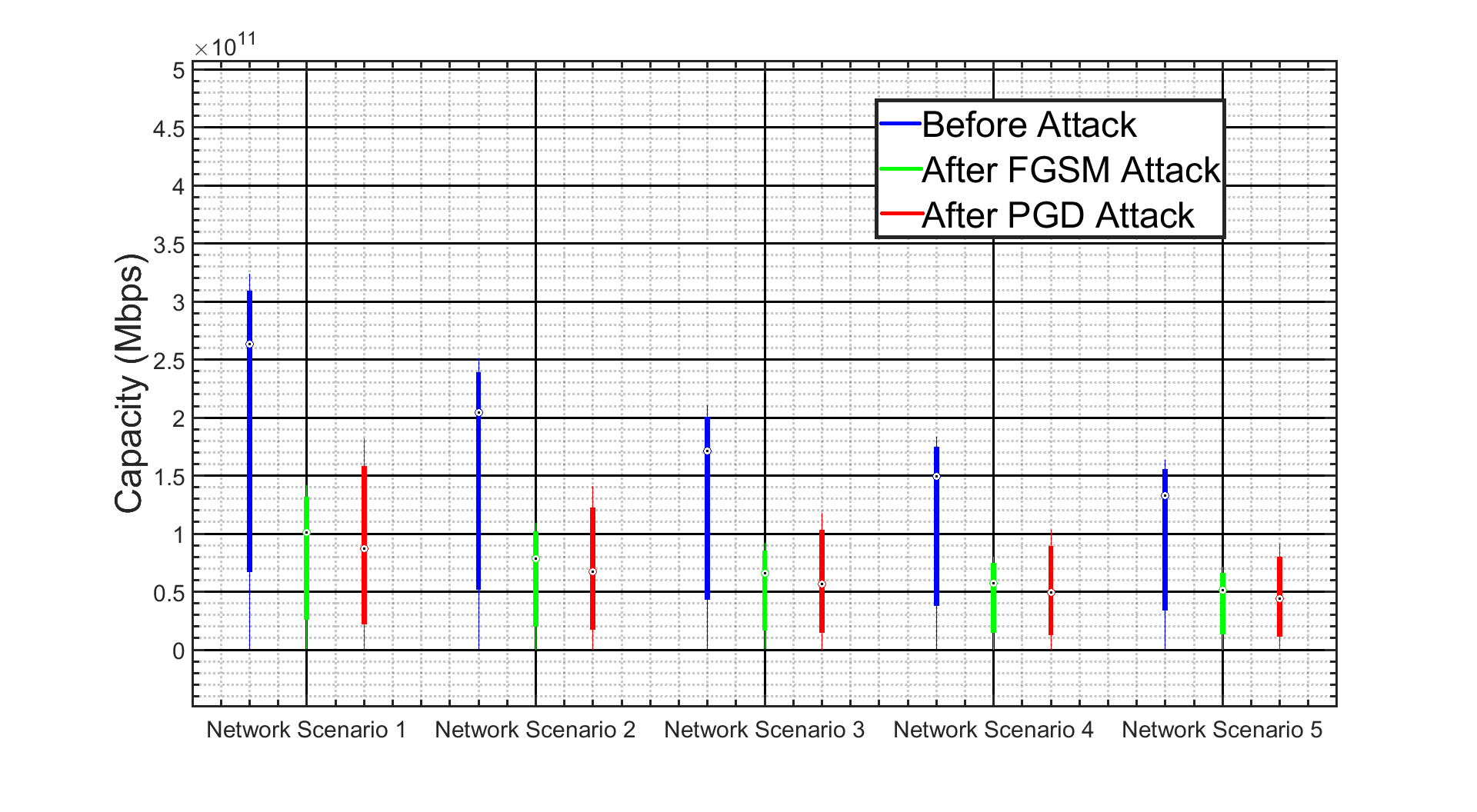}
\caption{The range of capacity of different network scenarios before and after the attack.} \label{fig:attack performance capacity}


\end{figure}

Finally, we have measured our end-to-end framework with different network scenarios and their range of capacity before and after the attack in Fig. \ref{fig:attack performance capacity}. We have simulated five distinct network scenarios described in Table \ref{tab: scenarios} where we observe a much lower capacity range after the FGSM and PGD attacks compared to the capacity range before the attack.

\begin{table}[t]
\centering
\caption{Different network scenarios.}
\begin{tabular}{|c|c|c|}
\cline{1-3}
 &  SNR [dB] & Number of users     \\ \cline{1-3}
Network Scenario 1& 25 & 50     \\ \cline{1-3}
Network Scenario 2 &  20& 40   \\ \cline{1-3}
Network Scenario 3 & 15 &  30    \\ \cline{1-3}
Network Scenario 4 & 10 & 20     \\ \cline{1-3}
Network Scenario 5 & 5 &  10    \\ \cline{1-3}
\end{tabular}\label{tab: scenarios}
\end{table}







\section{Conclusion}
In this paper, we assessed the vulnerability of the ML-based xApp at the near-RT RIC in O-RAN against adversarial attacks. As an example xApp, we deployed the interference classifier xApp to classify the interference type to mitigate the interference. In the meanwhile, there exists a malicious xApp that aims to perturb the original data that is stored in the database to cause misclassification at the interference classifier xApp. We considered two methods to perturb the stored data which are FGSM and PGD attacks. As a result of the adversarial attacks, we showed that the performance of the legitimate xApp is significantly affected which causes serious degradation of network performance.

\bibliographystyle{IEEEtran}
\bibliography{IEEEabrv,references}

\end{document}